\documentstyle{aa}     
\input psfig.sty   
\newcommand{\asec}  {\hbox{$^{\prime\prime}$}}
\newcommand{\amin}  {\hbox{$^{\prime}$}}
\begin{document}
\topmargin=+0.3cm
\thesaurus{ 
	           09.07.1;  
            09.01.1;  
            09.13.2;  
            11.09.4;  
            11.17.1;  
            13.19.3;  
	         }
   \title{Search for LiH in the ISM towards B0218+357}
   \subtitle{}
   \author{F.~Combes\inst{1} and T.~Wiklind\inst{2}}
   \offprints{F.~Combes, bottaro@obspm.fr}
   \institute{DEMIRM, Observatoire de Paris, 61 Av. de l'Observatoire,
              F--75014 Paris, France
   \and
Onsala Space Observatory, S--43992 Onsala, Sweden}
   \date{Received date; Accepted date}
   \maketitle
%
   \begin{abstract}
We report a tentative detection with the IRAM 30m telescope of the 
LiH molecule in absorption in front of the lensed quasar B0218+357. 
We have searched for the $J = 0 \rightarrow 1$ 
rotational line of lithium hydride
at 444 GHz (redshifted to 263 GHz). The line, if detected, is
optically thin, very narrow, and corresponds to a column density
of N(LiH) = 1.6 10$^{12}$ cm$^{-2}$ for an assumed excitation temperature
of 15 K, or a relative abundance
LiH/H$_2 \sim$ 3 10$^{-12}$. We discuss the implications of this
result.
   \end{abstract}
\section{Introduction}

Primordial molecules are thought to play a fundamental role in the 
early Universe, when stellar nucleosynthesis has not yet enriched 
the interstellar medium. After the decoupling
of matter and radiation, the molecular radiative processes, and 
the formation of H$_2$, HD and LiH contribute significantly 
to the thermal evolution of the medium (e.g. Puy et al 1993, Haiman,
Rees \& Loeb 1996). Even at the present time, it would be 
essential to detect such primordial molecules, to trace H$_2$ 
in the low-metallicity regions (e.g. Pfenniger \& Combes 1994, Combes
\& Pfenniger 1997). Unfortunately, the first transition of HD is
at very high frequency (2.7 THz), and the first LiH line, although
only at 444 GHz, is not accessible from the ground at $z=0$ due
to H$_2$O atmospheric absorption. This has to wait the launching of a
submillimeter satellite. 

Although the Li abundance is low (10$^{-10}$-10$^{-9}$), the observation
of the LiH molecule in the cold interstellar medium looks promising, 
because it has a large dipole moment, $\mu = 5.9$ Debye
(Lawrence et al.~1963), and the first rotational level is at
$\approx 21\,\rm K$ above the ground level, the corresponding
wavelength is $0.67\,\rm mm$ (Pearson \& Gordy 1969; Rothstein
1969). The line frequencies in the submillimeter and far-infrared
domain have been recently determined with high precision
in the laboratory (Plummer et al 1984, Bellini et al 1994).
Because of the great astrophysical interest of this molecule 
(e.g. Puy et al 1993), an attempt has been made to detect
LiH at very high redshifts ($z \sim 200$) 
with the IRAM 30m telescope (de Bernardis et al 1993).
 It has been proposed that the LiH molecules could smooth the primary
CBR (Cosmic Background Radiation)
anisotropies, due to resonant scattering, or create secondary
anisotropies, and they could be the best way to detect primordial
clouds as they turn-around from expansion (Maoli et al 1996,
but see also
Stancil et al 1996, Bougleux \& Galli 1997).

There has recently been some controversy about
the abundance of LiH. The computations of Lepp \& Shull (1984)
estimated the LiH/H$_2$ abundance ratio in primordial diffuse clouds
to be as high as 10$^{-6.5}$. With H$_2$/H $\sim$ 10$^{-6}$, the
primordial LiH/H ratio is $\sim$10$^{-12.5}$.
More recently, Stancil et al. (1996) computed an LiH/H abundance of 
$< 10^{-15}$ in the postrecombination epoch, since quantum mechanical
computations now predict
the rate coefficient for LiH formation through radiative association 
to be 3 orders of magnitude smaller than previously thought
from semi-classical methods (Dalgarno et al 1996).
In very dense clouds, however, three-body association reactions must
be  taken into account, and a significant fraction of all lithium will 
turn into molecules. Complete conversion due to this process requires
  gas densities of the order $\sim 10^9$\,cm$^{-3}$, rarely found in
  the general ISM. However, taken other processes into account, such
  as dust grain formation, an upper limit to the LiH abundance is
  the complete conversion of all Li into molecular form, with
  LiH/H$_2$ $\la 10^{-10}-10^{-9}$. With a LiH column density of
  10$^{12}$\,cm$^{-2}$, or N(H$_2$)$= 10^{22}$\,cm$^{-2}$, the optical
  depth of the LiH line will reach $\sim$1, in cold clouds
of velocity dispersion of $2\,\rm km\,s^{-1}$. The line should then be
easily detectable in dense dark clouds in the present interstellar
medium (like Orion where the column density reaches 10$^{23}$-10$^{24}$
cm$^{-2}$). This is a
fundamental step to understand the LiH molecule formation,
in order to interpret future results on primordial clouds,
although the primordial abundance of Li could be increased
by about a factor 10 in stellar nucleosynthesis (e.g. Reeves 1994).
 Once the Li abundance is known as a function of redshift, it
could be possible to derive its true primordial abundance,
a key factor to test Big Bang nucleosynthesis (either homogeneous or not).

Up to now, due to atmospheric opacity, no astrophysical LiH line 
has been detected, and the abundance of LiH in the ISM is unknown.
The atmosphere would allow to detect the isotopic molecule LiD
(its fundamental rotational line is at 251 GHz), but 
it has not been seen because
of the low D/H ratio, and the expected insufficient optical depth of LiH
\footnote{the LiD line at 251 GHz is not covered in the 247-263 GHz survey 
of Orion by Blake et al 1986,
but was observed at the McDonald 5m-telescope, Texas, see
Lovas 1992; we have ourselves checked with the SEST telescope that no
line is detected towards Sagittarius-B2 at this frequency.
The 3$\sigma$ upper limit to the LiD abundance towards SgrB2 is
$1 \times 10^{11}$\,cm$^{-2}$.}.

Another method to avoid atmospheric absorption lines is to observe 
a remote object, for which the lines are redshifted into an
atmospheric window.
 Here we report about the first absorption search for a LiH line at 
high redshift: the latter allows us to overcome the earth atmosphere
opacity, and thanks to the absorption technique we benefit of an excellent
spatial resolution, equal to the angular size of the B0218+357 quasar
core, of the order of 1 milli-arcsec (Patnaik et al 1995). At the
distance of the absorber (redshift $z=0.68466$, giving an angular size 
distance of 1089 Mpc, for $H_0$=75 km/s/Mpc and $q_0$ =0.5), this corresponds
to 5pc. We expect a detectable LiH signal, since the H$_2$ column
density is estimated to be N(H$_2$)$ = 5 \times 10^{23}$ cm$^{-2}$.
Menten \& Reid (1996) derive an N(H$_2$) value ten times lower than this, 
using the H$_2$CO($2_{11}-2_{12}$) transition at 8.6\,GHz. At this 
low frequency the structure and extent of the background continuum
source may be quite larger than at 100--200 GHz and the source
covering factor smaller. This means that their estimate
of the column density is a lower limit.

\section { Observations }

The observations were made with
the IRAM 30m telescope at Pico Veleta near Granada, Spain.
They were carried out in four observing runs, in December 1996, March,
July and December 1997.
Table 1 displays the observational parameters. 
We observed at 263 GHz with an SiS receiver tuned in single sideband (SSB).
The SSB receiver temperature varied between 400 and 450K, 
the system temperature was 600-1400K depending on weather conditions,
and the sideband rejection ratio was 10dB
(the image frequency is at 271.5 GHz, 
in a region where the atmospheric opacity increases rapidly due
to water vapour).
We used a 512x1MHz filterbank and an autocorelator 
backend, with 0.3 km/s resolution. 
We present here only the 1MHz resolution spectra, smoothed
to a 2.3 km/s channels, to improve the signal to noise.

\begin{table}
\begin{flushleft}
\caption[]{ Parameters for the tentative LiH line }
\begin{tabular}{lccl}
\hline
\multicolumn{1}{l}{J$_u$--J$_l$ }                           &
\multicolumn{1}{c}{1--0 }              &
\multicolumn{1}{c}{}                   \\
$\nu_{lab}$ GHz	& 443.953 & \\
$\nu_{obs}$ GHz	& 263.527 & \\
 Forward eff.	  &  0.86 &	\\
 Beam eff.	     &  0.32 & \\
 T$_A^*$        & 7 mK	& depth of absorption line \\
 T$_{\rm cont}$ & 15 mK & \\ 
 FWHM      	    & 3.2	km/s & \\
 $\sigma$       & 1.8 mK	& noise rms with $\Delta v$ 2.3 km/s \\
\hline
\end{tabular}
\, \\
\vskip 2truemm
$\alpha$(1950) = 02h 18m 04.1s  \\          
$\delta$(1950) = 35$^\circ$ 42\amin \, 32\asec  \\          
\end{flushleft}
\end{table}

The observations were done using a nutating subreflector with a 1' beamthrow
in azimuth.
We calibrated the temperature scale every 10 minutes by a chopper wheel on
an ambient temperature load, and on liquid nitrogen. Pointing was checked on
broadband continuum sources, and was accurate to 3\asec \, rms.
The frequency tuning and sideband rejection ratios were checked
by observing molecular lines towards Orion, DR21 and IRC+10216.

We integrated in total for 85 hours on the 263 GHz line,
and obtained a noise rms level of 1.8 mK in the T$_A^*$ antenna temperature
scale, with a velocity resolution of 2.3 km/s. The forward and beam 
efficiencies at the observed frequency are displayed in Table 1.
The continuum level was estimated by observing in a rapid on--off mode
using a special continuum backend. The switch frequency of the subreflector
was increased from 0.5 Hz to 2 Hz.

\begin{figure}
\psfig{figure=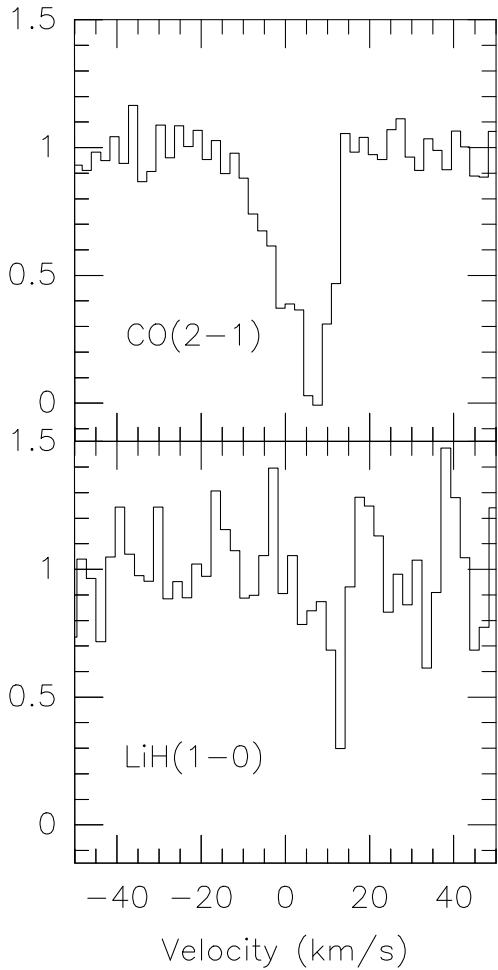,bbllx=3cm,bblly=65mm,bburx=11cm,bbury=195mm,width=8cm}
\caption[]{ Spectrum of LiH in its fundamental line (1--0)
at 444 GHz, redshifted at 263 GHz, in absorption towards B0218+357,
compared to the highly optically thick CO(2--1) line
previously detected.
The tentative LiH line is slightly shifted from the center by
about 5km/s, but is still comprised within the CO(2--1) velocity range.
Its width is compatible with what is expected from
an optically thin line.
Spectra have been normalised to the absorbed continuum level and
the velocity resolution is 2.3 km/s }
\label{lih_f1}
\end{figure}

\section{The molecular asborption line system towards B0218+357}

We select the ISM in front of the B0218+357 BL Lac object, because
it revealed the highest molecular column densities in all cases
of molecular absorption at high redshift
(Wiklind \& Combes 1995, Combes \& Wiklind 1996). The remote quasar
is gravitationally lensed by a foreground  galaxy at z=0.68466, which
produces the absorption. The radio image of the quasar is composed of two
distinct flat-spectrum cores (A and B component),  with a small 
 Einstein ring surrounding the B image, of 335 milli-arcsecond (mas) in 
diameter (Patnaik et al 1993). Since the ring has a steeper spectrum, it is 
interpreted as the image of a jet component, or in fact a hot spot or knot in 
the jet that happens to be just in the line of sight of the lens center. 

The intensity ratio between the two images is A/B $\approx$ 3.3 at several 
radio wavelengths, and it can vary slightly
(the B-component has varied in flux by $\approx$ 
10\% in a few months, O'Dea et al. 1992, Patnaik et al. 1993).
A large variety of molecules has been detected in absorption
towards B0218+357, among them several of the isotopes of CO, HCN, HCO$^+$, 
HNC, H$_2$O etc.. (Wiklind \& Combes 1995, Combes \& Wiklind 1995, 1997).
 Since the depth of the molecular absorption is less than
the continuum level, while being optically thick, we deduced that
the absorbing material does not cover the whole surface of the continuum.
This has been directly checked through high-resolution
interferometry (Menten \& Reid 1996, Wiklind \& Combes 1998).
 Only the A image is covered by
molecular clouds, and the fraction of the total continuum which 
is absorbed is $\sim$ 70\%.

Using the Jet Propulsion Laboratory catalog of molecular transitions
(Poynter \& Pickett 1985), we have checked whether the observed
absorption line could be caused by another molecule. This was done
for both signal and image frequencies (263.5 GHz and
271.5 GHz) and for z$=$0 and 0.68466. At z$=$0, we have
also looked for absorption of Galactic molecular gas using the
HCO$^+$(1--0) line at 89 GHz, without seeing any hint of absorption.

\section{Results and discussion}

Figure \ref{lih_f1} presents our LiH spectrum, compared to
that of CO(2--1) previously detected with the IRAM
30m-telescope (Wiklind \& Combes 1995, Combes \& Wiklind 1995).
There is only a tentative detection of LiH at $\sim$ 3 $\sigma$.
The line is very narrow, but is compatible to what is expected
from an optically thin line.
The CO(2--1) is highly optically thick, with $\tau \sim$ 1500. This
optical depth is determined from the detection of C$^{18}$O(2--1),
which is moderately thick, and the non--detection of C$^{17}$O(2--1).
 The center of the tentative line is shifted by 5 km/s from the
average center of other lines detected towards B0218+357. This shift
cannot be attributed to uncertainties of the line frequency, since
it has been measured in the laboratory (e.g. Bellini et al 1994),
and the error is at most 0.24 km/s at 3$\sigma$, once redshifted.
 But the scatter of the line centers is $\sim$ 3 km/s, and the width
of most of the lines is $\sim$ 15 km/s (cf Wiklind \& Combes 1998). The 
velocity shift is therefore insufficient to reject the line as real.

Combining our own continuum data with that of lower frequencies (obtained
from the NASA Extragalactic Database NED), we have previously found
that the continuum spectra of B0218+357 can be fitted with a power
law of slope --0.25 (Combes \& Wiklind 1997). This would imply a
continuum level of 15.5 mK at 263 GHz, which is in accord with the
measured level. Since only 70\% of the continuum is covered by
molecular gas, the continuum level to be used for our LiH observations
amounts to 11 mK.

\smallskip

We can write the general formula, concerning the total column density of 
the LiH molecule, 
observed in absorption between the levels $l \rightarrow u$ with an
optical depth $\tau$ at the center of the observed line of width $\Delta v$
at half-power:
$$
N_{LiH} = {{8\pi}\over{c^3}} f(T_x) {{\nu^3 \tau \Delta v} \over {g_u A_u} }
$$
where $\nu$ is the frequency of the 
transition, $g_u$ the statistical weight of the upper level
($= 2 J_u+1$), $A_u$ the Einstein coefficient of the transition, 
$T_x$ the excitation temperature, and 
$$
f(T_x) = {{Q(T_x) exp(E_l/kT_x)} \over { 1 - exp(-h\nu/kT_x)}}
$$
where $Q(T_x)$ is the partition function.
For the sake of simplicity, we adopt the hypothesis of restricted 
Thermodynamical Equilibrium conditions, i.e. that the excitation
temperature is the same for all the LiH ladder.
Since the line is not optically thick, but the optical thickness reaches 
$\tau$ = 1.3 at the center of the line, 
we have derived directly from the spectrum, through a Gaussian
fit of the opacity, the integrated $\tau \Delta v$  = 3.64 km/s. 
From the formulae above, and assuming an excitation temperature of
$T_x$ = 15 K (see Table 2 for variation of this quantity), we derive
a total LiH column density of 1.6 10$^{12}$ cm$^{-2}$ towards
B0218+357. Compared to our previously derived H$_2$ column
density of 5 10$^{23}$  cm$^{-2}$,
this gives a relative abundance of LiH/H$_2$ $\sim$ 3 10$^{-12}$.
Note that there is a possible systematic uncertainty associated with
this measure, due to the velocity difference between the maximum
opacity of the CO, HCO$^+$ and other lines with that of LiH.

\begin{table}
\begin{flushleft}
\caption[]{Derived LiH column density }
\begin{tabular}{lccccc}
\hline
 & & & & & \\
\multicolumn{1}{c}{$T_x$ }                     &
\multicolumn{1}{c}{(K)}                   &
\multicolumn{1}{c}{5 }                   &
\multicolumn{1}{c}{10 }                   &
\multicolumn{1}{c}{15 }                   &
\multicolumn{1}{c}{20 }                  \\
 & & & & & \\
\hline
 & & & & & \\
N(LiH) & (10$^{12}$ cm$^{-2}$)  & 0.4 & 0.9 & 1.6 & 2.4 \\
 & & & & & \\
LiH/H$_2$ & (10$^{-12}$) & 0.8  & 1.8 & 3.2 &  5 \\
 & & & & & \\
\hline
\end{tabular}
\end{flushleft}
\end{table}

\smallskip

To interpret this result, comparison should be made with the atomic 
species. First, it is likely that the molecular cloud on the line of
sight is dense and dark, and all the hydrogen is molecular, f(H$_2$) = 0.5.
The Li abundance (main isotope $^7$Li) at $z=0.68466$ 
(i.e 5-10 Gyr ago) can be estimated 
at Li/H  $\sim 10^{-9}$, since its abundance in the ISM increases with time.
The primordial Li abundance must be similar to that in metal deficient unevolved 
Population II stars, Li/H = 1-2 10$^{-10}$ (Spite \& Spite 1982), but
Li could be depleted at the stellar surface by internal mixing.
In meteorites and unevolved, unmixed Pop I stars, Li/H $\sim 10^{-9}$,
representative of the Li abundance some 4 Gyr ago. The present abundance
in the ISM is estimated around 3 10$^{-9}$ (Lemoine et al 1993).

We therefore deduce LiH/Li $\sim$ 1.5 10$^{-3}$. 
The uncertainty associated with the derived abundances are
large, but the low LiH/Li ratio seems to exclude complete transformation
of Li into LiH, as would be
expected in very dense clouds (e.g. Stancil et al 1996,
although the Li chemistry is not yet completely understood
in dark clouds).
However, it is likely that the cloud is clumpy, and in some of the more
diffuse parts, LiH is photodissociated (e.g. Kirby \& Dalgarno 1978).
Also, some regions of the cloud could have a
higher excitation temperature, in which case our computation
under-estimates the LiH abundance (although the absorption
technique selects preferentially cold gas, and the black-body temperature
at the redshift of the absorbing molecules is $T_{bg}$ = 4.6 K).

The present observations suggest that the detection of LiH
in emission towards dense clouds in the Milky Way should be
easy with a submillimeter satellite, provided that the 
spatial resolution is enough to avoid dilution of the
dense clumps.
It is also interesting to observe the rarer molecule $^6$LiH, which 
in some clouds might be of same order of abundance as the main 
isotopic species. Through optical absorption lines
Lemoine et al (1995) find towards two
velocity components in $\zeta$-Oph, $^7$Li/$^6$Li = 8.6 and 1.4.  Since 
$^6$Li is formed only in negligible amounts in the Big Bang, this 
ratio indicates that cosmic ray spallation has increased significantly
the Li abundances.

\vspace{0.25cm}

\acknowledgements
This work could not have been done without the 
 generous support from the IRAM-30m staff.
We also thank the referee, Daniele Galli, for useful
and interesting comments. Bibliographic and photometric data have 
been retrieved in the NED data base.
TW acknowledges financial support from the Swedish Natural Science
Council (NFR).

\end{document}